# One-Time Pads from the Digits of Pi

Devlin Gualtieri
Tikalon LLC, Ledgewood, New Jersey
(gualtieri@ieee.org)

*I present a method for generating one-time pads from the digits of pi. Computer code is given to generate such pads from passphrases in a method having an extremely low probability ($<10^{-53}$) of a successful discovery of the one-time pads by a brute-force attack. The advantages and disadvantages of this method are discussed.*

Introduction

The one-time pad is a secure encryption method when the pads are generated from a true source of random numbers and are kept private between the senders are receivers. There are presently available numerous physical sources capable of generating high quality random numbers at high rates. While generating one-time pads is easy, the bottleneck in their usage is their secure transmittal between sender and receiver. It is far easier for a sender and receiver to decide on a method to retrieve passphrases at planned locations on Internet web sites or in a physical document, such as a book.

Digits of Pi

The mathematical constant, pi, has been computed to tens of trillions of digits through the aid of rapidly convergent series. While the digits of pi have not been proven to be normally distributed, statistical tests show that they are random to a high degree.

An interesting development in the computation of digits of pi is the Bailey–Borwein–Plouffe (BBP) formula that allows computation of the nth hexadecimal digit of pi without computation of its preceding hexadecimal digits.[1] This formula allows selection of a range of supposedly random digits from an extremely large pool of random numbers.

One-Time Pads

Creating a one-time pad from a range of digits of pi can be done by a simple computation, but such a pad can be discovered by a brute-force search of the digits of pi. A desktop computer might contain a hard drive with a 500 gigabyte capacity from which a range of random numbers from a trillion hexadecimal digits of pi can be selected. However, it would take just a little more time for another computer to search through these digits to discover the one-time pad. Also, an application-specific integrated circuit (ASIC) could be developed to execute this search quite quickly when the digits of pi are stored in solid state memory.

Adding Complexity

When two one-time pads are combined through an exclusive-or (**XOR**) operation, the result is a new one-time pad. We can select **m** ranges of digits of pi and combine these using **XOR** to produce a new one-time pad. Using just the first million digits of pi (plus the size of the pad, which is much smaller than a million) as our random number pool with ten **XOR** operations gives us a large number of unique one-time pads. The combination of a million things taken 10 at a time is given as

$$_{1000000}C_{10} = (1000000!)/(10!(999990!)) = 2.75 \times 10^{53}$$

This is an extreme number of possible combinations that leads to a low probability of success of a brute-force attack. The problem then becomes one of how to select these **m** ranges from the digits of pi.

Passphrases and Hashing

SHA-2 (Secure Hash Algorithm 2) is used as the hashing algorithm in my simple example. SHA-256 produces a 256-bit hash (64 hexadecimal digits) that can be partitioned easily into ten hexadecimal numbers of length five that cover a range of 1,048,576 elements. These five numbers, $a_i$, are used as the starting address for 256 byte random number blocks in the hexadecimal representation of pi. The digits of pi in the example code, shown in *Appendix I*, are computed using the May, 4, 1996, program created by David H. Bailey.[2] A visual representation of the method is shown in fig. 1.

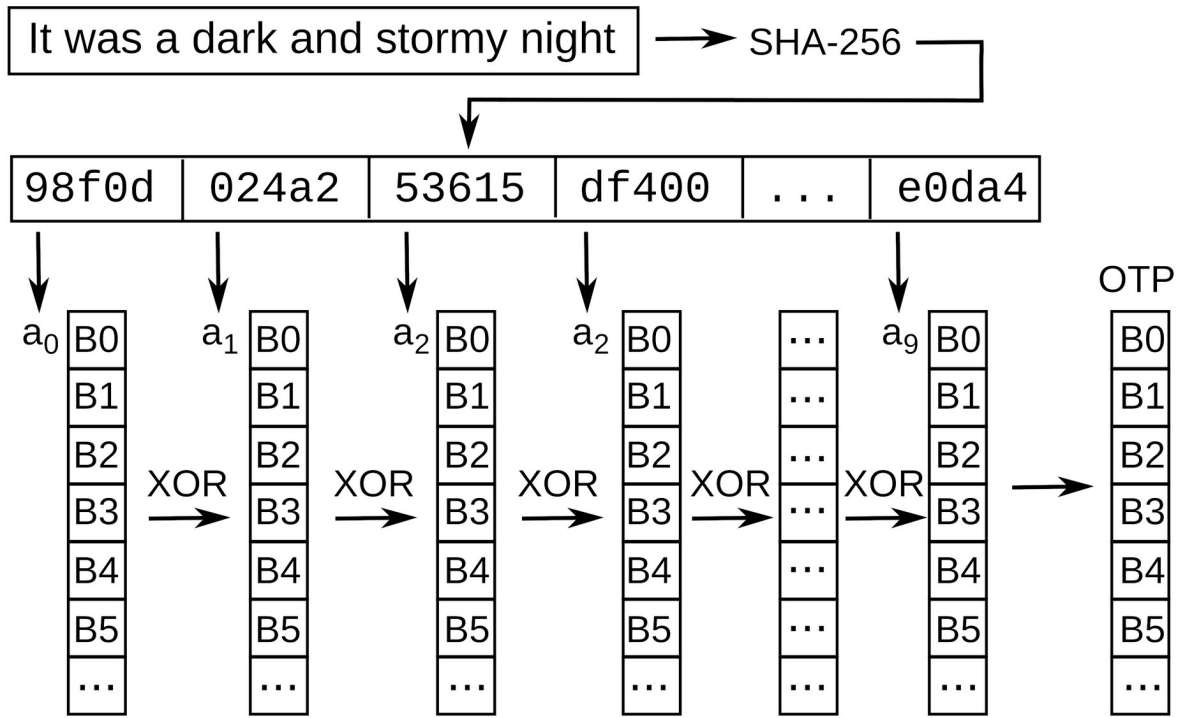

Fig. 1 Transformation of a hashed passphrase into data addresses $a_0$-$a_9$ for blocks of digital bytes $B_0$-$B_n$ of pi. The blocks are bitwise XORed to produce the one-time pad.

The outdated Data Encryption Standard (DES) was rendered more secure by applying the algorithm three times using three different passphrases as Triple DES (3DES). I've taken the same approach as an option in the example code.

While the seemingly random digits of pi are used in this method, a large file of random numbers created from a high-quality random number source can be used, also. The disadvantage of this approach is that this file must be shared by communicators, although this one-time sharing allows generation of a multiplicity of one-time pads.

Passphrase Distribution

This method eliminates the problem of secure distribution of one-time pads, but it introduces the similar problem of passphrase distribution. In this case, we appeal to a steganographic technique of communicators extracting passphrases from copies of a common document, such as a book. The book can be a physical copy held at each communicating site, or an electronic copy available on the Internet.

As an example based on our pi motif, a passphrase can be derived from a phrase starting with the 31st word on page 41 of the PDF file of Plato's Republic, available at the Internet Archive (*be firs a husbandman*, which gives an SHA-256 hash of 61d44985f82b046740d3ac4f0c0e291ffe6bf3bc6fe4a3d5169fb7523f178d9a).[3]  The security of such passphrases depends on the identity of the document and the shared algorithm for selection of elements within the document both being private.  The elements can also be prearranged byte locations in a digital document, video or audio file.

Statistical Tests

The example code of *Appendix I* was too inefficient to generate enough data for the Dieharder statistical tests.[4]  However, it was possible to generate 500,000 random bytes by concatenating the one-time pads of repeated executions of the example program using as passphrases the concatenation of a numerical timestamp and a GNU C random number implemented as a linear feedback shift register.  An example of such a passphrase is 202103142347 concatenated with 11767 to give 20210314234711767.  Fig. 2 is a histogram of the occurrence of byte values for these 500,000 bytes.

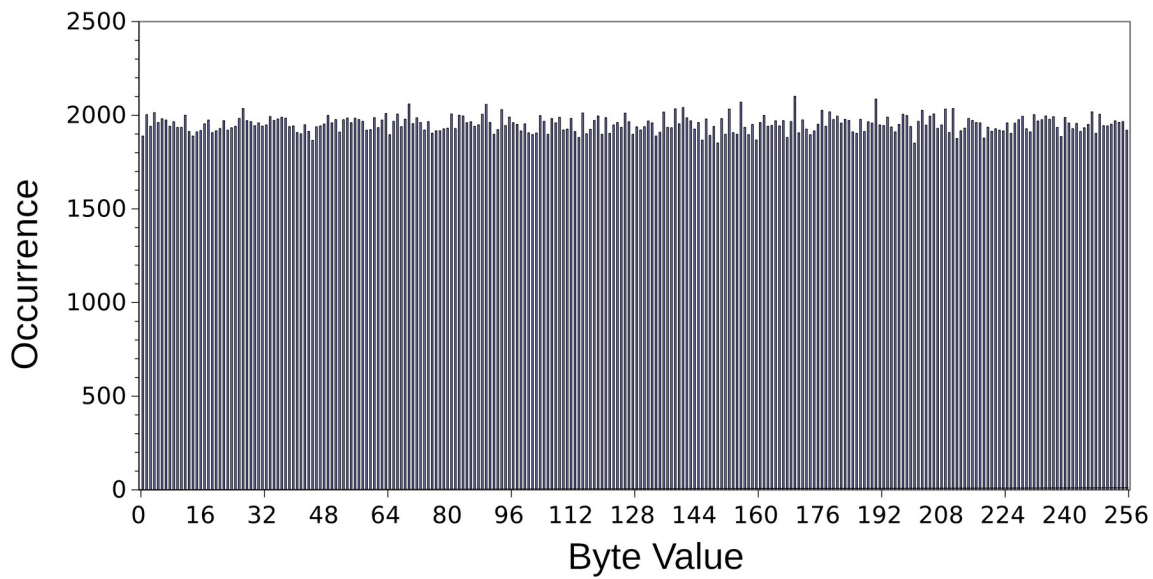

Fig. 2  Histogram of the occurrence of byte values for 500000 generated random bytes.

The mean of these values is 1953.125, and the standard deviation is 41.893,  The coefficient of variation, also known as relative standard deviation, is 2.1%.  It's noted that the value of $(1/\sqrt{1953.124})$ is 0.0226.

Discussion

There are several obvious disadvantages of this approach, as follow:

1)  The randomness of the digits of pi has not been proven.
2)  Calculating digits of pi at individual starting addresses is computationally inefficient. Speed would be increased by reading from a file.
3)  A usable method of passphrase distribution, possibly by steganography. must be established prior to pad generation and usage.

The advantages of this approach are that the one-time pads themselves need not be distributed and the number of pads that can be constructed is large.

Appendix I – Source code for generating one-time pads from the digits of pi.

```c
/* -*- Mode: C; indent-tabs-mode: t; c-basic-offset: 4; tab-width: 4 -*- */
/*
 * pi_OTP.c
 * Copyright (C) dmgualtieri 2021 <gualtieri@ieee.org>
 *
 * pi_OTP is free software: you can redistribute it and/or modify it
 * under the terms of the GNU General Public License as published by the
 * Free Software Foundation, either version 3 of the License, or
 * (at your option) any later version.
 *
 * pi_OTP is distributed in the hope that it will be useful, but
 * WITHOUT ANY WARRANTY; without even the implied warranty of
 * MERCHANTABILITY or FITNESS FOR A PARTICULAR PURPOSE.
 * See the GNU General Public License for more details.
 *
 * You should have received a copy of the GNU General Public License along
 * with this program.  If not, see <http://www.gnu.org/licenses/>.
 */

/*  Pi calculation adapted from David H. Bailey code of 960429
    Bailey notes that this code is valid up to ic = 2^24 on systems with IEEE arithmetic.
    This version 13 February 2021
      To compile: gcc pi_OTP.c -o pi_OTP -lm
      Usage: ./pi_OTP "passphrase1" ["passphrase2" "passphrase3"]
*/

#include <stdio.h>
#include <strings.h>
#include <math.h>
#include <time.h>
#include <sys/types.h>

#define NHX 16
#define pad_size 256
#define COMMAND_LEN 80
#define DATA_SIZE 65

//Prototypes
void exit(int exit_code);
char *strcpy(char *dest, const char *src);
char *strcat(char *dest, const char *src);
char *strncat(char *dest, const char *src, size_t n);
long int strtol(const char *str, char **endptr, int base);
void *memcpy(void *dest, const void *src, size_t n);
int isdigit(int c);

double pid, s1, s2, s3, s4;
double series(int m, int n);
void ihex(double x, int m, char c[]);
int ic = 0;
int i = 0;
int j = 0;
```

```c
    int k = 0;
    int n = 0;
    int num;
    int rounds = 1;
    char *hash;
    char password[3][64] = { "" };

    char *end_ptr;
    char address_buffer[8] = "";
    long pi_address[10];
    int retval;
    int temp = 0;
    char char_buffer[2] = "";
    char str_buffer[1 + pad_size] = "";
    int OTP[pad_size] = { 0 };

    char chx[NHX];
    char fn1[64] = "pi_hex.txt";  // File of hex digits of pi
    char st_time[100];
    time_t OTP_time;
    struct tm *t;
    clock_t begin;
    clock_t end;
    FILE *outdata;

    void ihex(double x, int nhx, char chx[])
    /*  This returns, in chx, the first nhx hex digits of the fraction of x.
    */
    {
        int i;
        double y;
        char hx[] = "0123456789abcdef";

        y = fabs(x);

        for (i = 0; i < nhx; i++) {
          y = 16. * (y - floor(y));
          chx[i] = hx[(int) y];
        }
    }

    double series(int m, int ic)
    /*  This routine evaluates the series  sum_k 16^(ic-k)/(8*k+m)
        using the modular exponentiation technique. */
    {
        int k;
        double ak, p, s, t;
        double expm(double x, double y);
    #define eps 1e-17

        s = 0.;

    /*  Sum the series up to ic. */

        for (k = 0; k < ic; k++) {
          ak = 8 * k + m;
          p = ic - k;
```

```
          t = expm(p, ak);
          s = s + t / ak;
          s = s - (int) s;
       }

/*  Compute a few terms where k >= ic.  */

       for (k = ic; k <= ic + 100; k++) {
          ak = 8 * k + m;
          t = pow(16., (double) (ic - k)) / ak;
          if (t < eps)
             break;
          s = s + t;
          s = s - (int) s;
       }
       return s;
}

double expm(double p, double ak)
/*  expm = 16^p mod ak.  This routine uses the left-to-right binary
    exponentiation scheme.  It is valid for  ak <= 2^24.  */
{
    int i, j;
    double p1, pt, r;
#define ntp 25
    static double tp[ntp];
    static int tp1 = 0;

/*  If this is the first call to expm, fill the power of two table tp.  */

    if (tp1 == 0) {
       tp1 = 1;
       tp[0] = 1.;

       for (i = 1; i < ntp; i++)
          tp[i] = 2. * tp[i - 1];
    }

    if (ak == 1.)
       return 0.;

/*  Find the greatest power of two less than or equal to p.  */

    for (i = 0; i < ntp; i++)
       if (tp[i] > p)
          break;

    pt = tp[i - 1];
    p1 = p;
    r = 1.;

/*  Perform binary exponentiation algorithm modulo ak.  */

    for (j = 1; j <= i; j++) {
       if (p1 >= pt) {
          r = 16. * r;
          r = r - (int) (r / ak) * ak;
```

```c
            p1 = p1 - pt;
        }
        pt = 0.5 * pt;
        if (pt >= 1.) {
            r = r * r;
            r = r - (int) (r / ak) * ak;
        }
    }

    return r;
}

char *get_popen_data(char *password)
{
    //Get SHA-256 hash of password using Linux command line call
    //Test vector when password = 'password' is
    //5e884898da28047151d0e56f8dc6292773603d0d6aabbdd62a11ef721d1542d8
    FILE *pf;
    char command[COMMAND_LEN] = "";
    static char data[DATA_SIZE];

    strcat(command, "echo -n ");
    strcat(command, password);
    strcat(command, " | sha256sum");

    // Setup our pipe for reading and execute our command.
    pf = popen(command, "r");

    if (!pf) {
      fprintf(stderr, "Could not open pipe for output.\n");
      return 0;
    }
    // Grab data from process execution
    fgets(data, DATA_SIZE, pf);

    // Print grabbed data to the screen.
    //fprintf(stdout, "%s\n",data);

    if (pclose(pf) != 0)
      fprintf(stderr, " Error: Failed to close command stream \n");

    return (data);
}

void get_hex_digits(int start, int number)
{
    //ic is the hex digit position -- output begins at position ic + 1.
    //hex digits stored in str_buffer
    int j = 0;
    int ic;
    str_buffer[0] = '\0';
    for (ic = start; ic < (start + number); ic = ic + 10) {
      s1 = series(1, ic);
      s2 = series(4, ic);
      s3 = series(5, ic);
```

```c
        s4 = series(6, ic);
        pid = 4. * s1 - 2. * s2 - s3 - s4;
        pid = pid - (int) pid + 1.;
        ihex(pid, NHX, chx);
        strncat(str_buffer, chx, 11);
        j = j + 1;
        if (j == 8) {
            j = 0;
        }
      }
}

int main(int argc, char *argv[])
{

    if (argc > 1) {
//passwords given on command line
      if (argc == 2) {
//single round
          strcpy(password[0], argv[1]);
          rounds = 1;
      }
      if (argc == 3) {
//two rounds
          strcpy(password[0], argv[1]);
          strcpy(password[1], argv[2]);
          rounds = 2;
      }
      if (argc == 4) {
//three rounds
          strcpy(password[0], argv[1]);
          strcpy(password[1], argv[2]);
          strcpy(password[2], argv[3]);
          rounds = 3;
      }
      printf("Rounds = %d\t%s\t%s\t%s\n", rounds, password[0],
            password[1], password[2]);
    } else {
      printf("Enter number of rounds (1-3): ");
      retval = scanf("%d", &rounds);
      if ((!isdigit(rounds)) || (rounds < 1) || (rounds > 3)) {
          printf("Number of rounds out of bounds! Exiting.");
          return 0;
      }

      begin = clock();

      for (i = 1; i < (1 + rounds); i++) {
          printf("Enter password %d: ", i);
          retval = scanf("%s", password[i - 1]);
          printf("\npassword = %s\n", password[i - 1]);
          hash = get_popen_data(password[i - 1]);
// Print grabbed data to the screen.
          fprintf(stdout, "%s\n", hash);
      }
    }
```

```c
//build OTP filename
    time_t OTP_time = time(NULL);
    struct tm *t = localtime(&OTP_time);
    strftime(st_time, sizeof(st_time), "OTP_%Y%m%d-%H%M%S.txt", t);
    strcpy(fn1, st_time);

    printf("\nOutput file selected = %s\n", fn1);

    if ((outdata = fopen(fn1, "w")) == NULL) {
      printf("\nOutput file cannot be opened.\n");
      exit(1);
    }

    for (k = 0; k < rounds; k++) {
      hash = get_popen_data(password[k]);
//build pi digit addresses from hash

      for (i = 0; i < 10; i++) {
          memcpy(address_buffer, hash + (5 * i), 5);
//add null termination
          address_buffer[5] = '\0';
//need for speed - truncate to four hex digits for testing
//address_buffer[4]='\0';
//XOR combine hash with previous round OTP
//in first round, OTP is all zeros, so hash is unchanged
          pi_address[i] = strtol(address_buffer, &end_ptr, 16);
          temp =
            (4096 * OTP[5 * i]) + (1024 * OTP[1 + (5 * i)]) +
            (256 * OTP[2 + (5 * i)]) + (16 * OTP[3 + (5 * i)]) +
            (OTP[4 + (5 * i)]);
//bitwise XOR is the ^ operator
          pi_address[i] = (pi_address[i] ^ temp);
          printf("%s\t%ld\n", address_buffer, pi_address[i]);
      }

//XOR digits at the ten addresses
      for (n = 0; n < 10; n++) {
          get_hex_digits(pi_address[n], 256);

//printf ("\nstr_buffer = %s\n",str_buffer);
//bitwise XOR is the ^ operator
          for (i = 0; i < pad_size; i++) {
            char_buffer[0] = str_buffer[i];
            char_buffer[1] = '\0';
            num = (int) strtol(char_buffer, &end_ptr, 16);
            OTP[i] = OTP[i] ^ num;
          }
          printf("%d - %s\n", n, str_buffer);
      }

    }

//print OTP to file
    printf("\nOTP: ");
    for (i = 0; i < pad_size; i++) {
      printf("%x", OTP[i]);
      fprintf(outdata, "%x", OTP[i]);
```

```c
    }

    fclose(outdata);
    printf("\n");

    end = clock();

    printf("Computation time = %.3f seconds\n",
           (double) (end - begin) / CLOCKS_PER_SEC);

    return 1;
}
```